\documentclass[pre,showpacs,byrevtex,amsmath,amssymb,nofootinbib,preprint]{revtex4}

\usepackage{graphicx}
\usepackage{dcolumn}
\usepackage{bm}

\begin{document}
\preprint{PREPRINT}

\title[Short Title]{Erratum: Effect of Image Forces on Polyelectrolyte Adsorption at a Charged Surface}

\author{Ren\'e Messina}
\email{messina@thphy.uni-duesseldorf.de}
\affiliation
{Institut f\"ur Theoretische Physik II,
Heinrich-Heine-Universit\"at D\"usseldorf,
Universit\"atsstrasse 1,
D-40225 D\"usseldorf,
Germany}

\date{\today}

\maketitle

Equation (6) of this paper \cite{Messina_PRE_2004}, which will be referred to as Paper 1, 
is erroneous and leads to an overestimation of the effect of image forces.
The correct form of that equation containing the dielectric term discontinuity 
$\Delta_{\epsilon}$  reads
%
\begin{eqnarray}
\label{eq:img}
\beta U_{\rm Coul}^{({\rm plate})}(z_i) = 
l_B\left[ \pm 2\pi \sigma_0 (1+\Delta_{\epsilon}) z_i +  \frac{\Delta_{\epsilon}}{4z_i}\right].  
\end{eqnarray}
%
It is precisely the term $(1+\Delta_{\epsilon})$ that was missing in Paper 1.
Despite of that error, most of our conclusions remain qualitatively correct.
The only conclusion that is truly affected concerns the overcharging.
With our new data based on the correct Eq. \eqref{eq:img}, it is
found that {\it overcharging is robust against image forces}.
Furthermore, all our results obtained in the absence of image charges 
(i.e., $\Delta_{\epsilon}=0$) in Paper 1 are evidently unaffected by this 
mistake.

\begin{figure}[b]
\includegraphics[width = 8.0 cm]{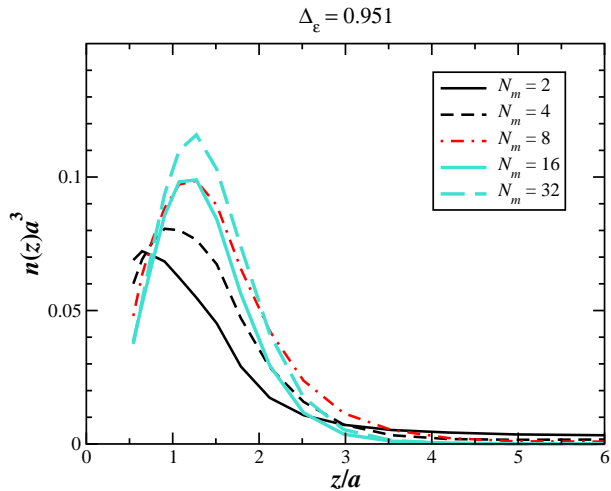}
\caption{
Profiles of the monomer density $n(z)$ for different chain length $N_m$ 
with $\sigma_0L^2=64$ and $\Delta_{\epsilon}=0.951$.
}
\label{fig.nz_Qp-64}
\end{figure}

We now briefly discuss the impact of our corrections
by providing some representative corrected data.
The profiles of the monomer distribution $n(z)$ can be found in Fig. \ref{fig.nz_Qp-64}
that corresponds to our earlier Fig. 1(b) in Paper 1.
From Fig. \ref{fig.nz_Qp-64}, it can be seen that 
the same qualitative behavior is found as that sketched in Fig. 1(b)
from Paper 1. The height of the peaks in Fig. \ref{fig.nz_Qp-64}
are roughly twice as large as those found in Fig. 1(b) from Paper 1.
Some data from Fig. \ref{fig.nz_Qp-64} were also published elsewhere
[see Fig. 3(b) in Ref. \cite{Messina_JCP_2006}].

%
\begin{figure}
\includegraphics[width = 8.0 cm]{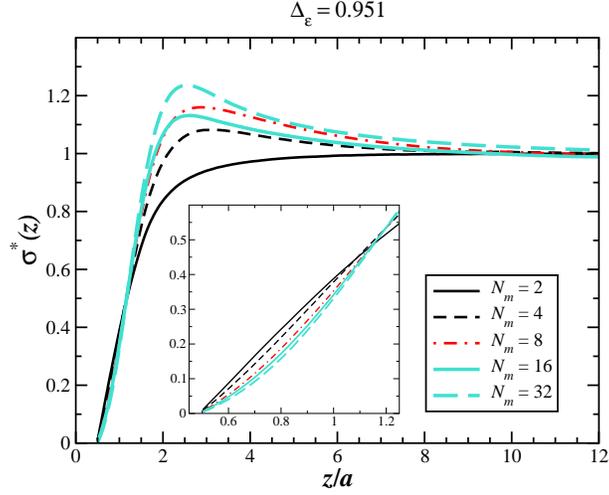}
\caption{
Profiles of the reduced net fluid charge $\sigma^*(z)$  for different chain length $N_m$ 
with $\sigma_0L^2=64$  with $\Delta_{\epsilon}=0.951$.
The inset is a magnification of the region near contact.
}
\label{fig.Qz_star_Qp-64}
\end{figure}
%
An important change concerns Fig. 4(b) in Paper 1 where
an erroneous cancellation of overcharging was reported.
The corrected data are now depicted in Fig. \ref{fig.Qz_star_Qp-64}.
The strength of the overcharging is presently nearly identical to
that obtained without image forces at $\Delta_{\epsilon}=0$ (compare with Fig. 4 from Paper 1).
Near contact ($z \lesssim 1.2a$)  it is found that the fraction of charge $\sigma^*(z)$ 
(see Fig. \ref{fig.Qz_star_Qp-64}), that is essentially due to the adsorbed monomers, 
decreases with growing $N_m$. This feature was already reported in Fig. 4(b) from Paper 1.



\end{document}